\documentclass[twocolumn,showpacs,preprintnumbers,amsmath,amssymb,letterpaper]{revtex4}
\usepackage{graphicx}
\usepackage{dcolumn}
\usepackage{bm}
\usepackage{amssymb}
\usepackage{epsfig}

\newcommand{\be}{\begin{equation}}
\newcommand{\ee}{\end{equation}}
\newcommand\beq{\begin{eqnarray}}
\newcommand\eeq{\end{eqnarray}} 
\newcommand\eqn[1]{\label{eq:#1}} 
\newcommand\eq[1]{eq. (\ref{eq:#1})}

\newcommand{\bfq}{{\mathbf q}}

\newcommand{\MeV}{{\rm ~MeV }}

\newcommand{\bfsigma}{\boldsymbol{ \sigma}}
\newcommand{\bftau}{\boldsymbol{ \tau}}
\begin{document}

\preprint{\hskip5.5in\vbox{CERN-PH-TH-2008-241\\ IFT-UAM/CSIC-08-89\\  INT-PUB-08-57\\ 
UNH-08-04\\ TUW-08-23}}

\title{Perturbative nuclear physics}
\author{Silas R. Beane$^{1}$}
\email{silas@physics.unh.edu}

\author{David B. Kaplan$^{2}$}
\email{dbkaplan@phys.washington.edu}

\author{Aleksi Vuorinen$^{3,4}$}
\email{aleksi.vuorinen@cern.ch}

\affiliation{$^1$ Department of Physics,
University of New Hampshire,
Durham, NH 03824-3568, USA }

\affiliation{$^2$Institute for Nuclear Theory, University of Washington, Seattle, WA 98195-1550, USA}
\affiliation{$^3$  CERN, Physics Department, TH Unit, CH-1211 Geneva 23, Switzerland}
\affiliation{$^4$  Institut f¬ur Theoretische Physik, TU Wien, Wiedner Hauptstr. 8-10, A-1040 Vienna, Austria   }

\begin{abstract}
We present a new formulation of effective field theory for
nucleon-nucleon (NN) interactions which treats pion interactions
perturbatively, and we offer evidence that the expansion converges
satisfactorily to third order in the expansion, which we have computed
analytically for $s$ and $d$ wave NN scattering.
Starting with the Kaplan-Savage-Wise (KSW) expansion about the
nontrivial fixed point corresponding to infinite NN scattering length,
we cure the convergence problems with that theory by summing to all
orders the singular short distance part of the pion tensor
interaction.  This method makes possible a host of high precision
analytic few-body calculations in nuclear physics.
 \end{abstract}
\date{\today}
\maketitle

\section{Introduction}

All strong interactions in nuclear physics are of finite range, and
therefore should be amenable to an effective field theory (EFT)
treatment at sufficiently low energy \cite{Beane:2000fx,Bedaque:2002mn,Kaplan:2005es}.  However,
in contrast to the Fermi EFT for the weak interactions, the strong
interactions between nucleons are nonperturbative even for momenta
much smaller than the inverse range of the interactions;  therefore
the effect of the leading four-fermion interaction must be treated to
all orders in perturbation theory, even though by conventional power
counting it is an ``irrelevant" operator.  Weinberg was the first to
describe an EFT for nuclear forces
\cite{Weinberg:1990rz,Weinberg:1991um,Weinberg:1992yk}, and devised
the prescription that one compute the nuclear potential in an EFT
expansion, truncate at a given order, and then solve the
Lippmann-Schwinger equation exactly with that potential.  This program
has since been pursued by a number of groups
\cite{Ordonez:1992xp,Ordonez:1993tn,vanKolck:1994yi,Ordonez:1995rz,Friar:1998zt,Rentmeester:1999vw,
Bernard:1996gq,Epelbaum:1999dj,Epelbaum:2000mx,Epelbaum:2002ji,Epelbaum:2003xx,Epelbaum:2004fk,
Entem:2001cg,Entem:2002sf,Entem:2003ft, 
PavonValderrama:2005wv,PavonValderrama:2005uj},
with very impressive fits to phase shift data at $\text{N}^3\text{LO}$.  An advantage of
this approach is that the long distance part of the interaction
correctly incorporates chiral symmetry; furthermore, with Weinberg's
power counting scheme for the EFT expansion, there is in principle a
systematic improvement of the results with increasing order.  A disadvantage of Weinberg's scheme is that it is not
renormalizable, in the sense that at any given order in the expansion
there are divergences that cannot be absorbed by operators included at
that order, arising from the singular nature of the EFT potential
\cite{Kaplan:1996xu,Beane:2001bc,Nogga:2005hy}.  Thus  results
depend on a regulator scale $\Lambda$ which cannot be removed, implying
that the treatment of short distance interactions is model-dependent;
in more recent developments the potential is regulated
separately from the Lippmann Schwinger equation, so that the result
depends on two independent regularization scales  \cite{Epelbaum:2003xx}.  An analysis of high
partial wave channels at NLO in the Weinberg EFT in
ref. \cite{Nogga:2005hy} demonstrated that the cutoff dependence was a
feature of all channels subject to an attractive pion tensor force ---
despite the fact that there is no local operator to absorb this model
dependence until order $(\ell+1)$ in the expansion for a channel with
angular momentum $\ell$.  Furthermore \cite{Nogga:2005hy} demonstrated that at this
order, observables in some channels (e.g. ${}^3P_0$) are particularly
sensitive to the value of the cutoff even at energies as low as
$T_\text{lab} = 50\MeV$. It is argued that predictions at a given
order only vary at the level of higher order corrections as the
regulator is varied over some range, so that the model dependence does
not interfere with the predictive power of the EFT.  This hope is
difficult to verify since the computations are all numerical, and the
numerical evidence suggests that the acceptable range for $\Lambda$ is
very narrow.

\begin{figure}[t]

\includegraphics[width=6.1cm]{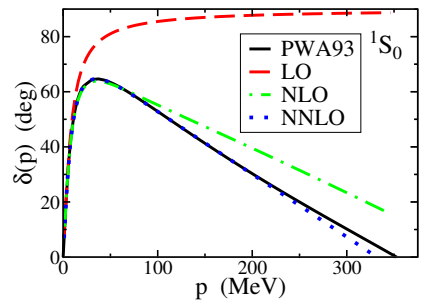}

\label{J0}

\caption[a]{The ${}^1S_0$ NN phase shift in the KSW expansion, versus
momentum in the center of mass frame to NNLO, compared with the Nijmegen
PWA93 partial wave analysis \cite{Stoks:1993tb}. Our calculation reproduces the result of
\cite{Fleming:1999bs,Fleming:1999ee}.}
\end{figure}

The alternative KSW theory entails an expansion of the NN scattering
amplitude, instead of the nuclear potential, effected by computing a
well-defined class of Feynman diagrams at each order in the expansion
\cite{Kaplan:1996xu, Kaplan:1998we,Kaplan:1998tg}.  KSW power counting is
not determined by how operators scale near the trivial IR fixed point
of the nucleon contact interaction (as in Fermi's weak
interaction EFT); instead it is determined by
operator scaling about the nontrivial UV fixed point corresponding to
infinite scattering length.  At this fixed point nucleon operators for
s-wave scattering develop large anomalous dimensions and are resummed
nonperturbatively, a reasonable starting point given how much larger
NN scattering lengths are than the range of their interaction. 

The KSW scheme expands the NN scattering amplitude in powers of $Q$,
where the nucleon momentum $p$, pion mass $m_\pi$ and the inverse scattering length $1/a$ are all
considered $O(Q)$, while other mass
scales such as the nucleon mass $M$, the pion decay constant $f_\pi$
are taken to be $O(1)$.  It was argued that convergence of the KSW expansion is governed by the scale
 $\Lambda_{NN} = 16\pi f_\pi^2/(g_A^2 M)= 300\MeV$.  
   An
advantage of this approach is that the scattering amplitudes can be
computed analytically, and at each order the amplitude is renormalized
and independent of the cutoff.  NN phase shifts were computed to order NNLO in refs. \cite{Fleming:1999bs,Fleming:1999ee}; the  result  for the
spin-singlet ${}^1S_0$ phase shift is shown in Fig.~1, plotted versus
the momentum $p$ of each nucleon in the center of mass frame.

Although successful in the spin-singlet channel, it was discovered in
ref.~\cite{Fleming:1999ee} that the KSW expansion does not converge in the
${}^3S_1$ channel, and the authors identified the singular tensor
potential mediated by pions, scaling as $-1/r^3$ for small $r$, to be the
cause of the failure.  Such a singular attractive interaction is incapable of
supporting a ground state and no contact interaction can
remedy this pathology. One possible solution suggested in \cite{Beane:2001bc} is to expand
around the chiral ($m_\pi=0$) limit, treating the infinite number of
bound states in the pion potential as being short-range and outside
the purview of the EFT.  In this letter we propose a different
solution: we modify the pion propagator in a manner reminiscent of
Pauli-Villars regulation characterized by a heavy mass scale
$\lambda$.  This modification tames the $1/r^3$ singularity in pion
exchange, effectively shifting that physics into the contact
interactions and reordering the summation of strong short-distance
effects.  The advantages of the KSW expansion are retained: there is a
well-defined power counting scheme that organizes the calculation, and
results are analytic.  Dependence on the scale $\lambda$ can therefore
be studied analytically, and we find that all contributions that grow
as powers of $\lambda$ are absorbed into counterterms. The limit
$\lambda\to \infty$ is therefore smooth, and the KSW expansion is
recovered in that limit. Here we present promising results for the
low-lying spin triplet phase shifts to NNLO that indicate convergence
of the expansion, and we discuss how the scale $\lambda$ resembles the
renormalization scale $\mu$ encountered in perturbative QCD
calculations: an unphysical scale which controls the ordering of the
perturbative expansion and its convergence.

\section{Short distance modification of the pion propagator}
\begin{figure*}[t]

\hbox{\includegraphics[width=6.1cm]{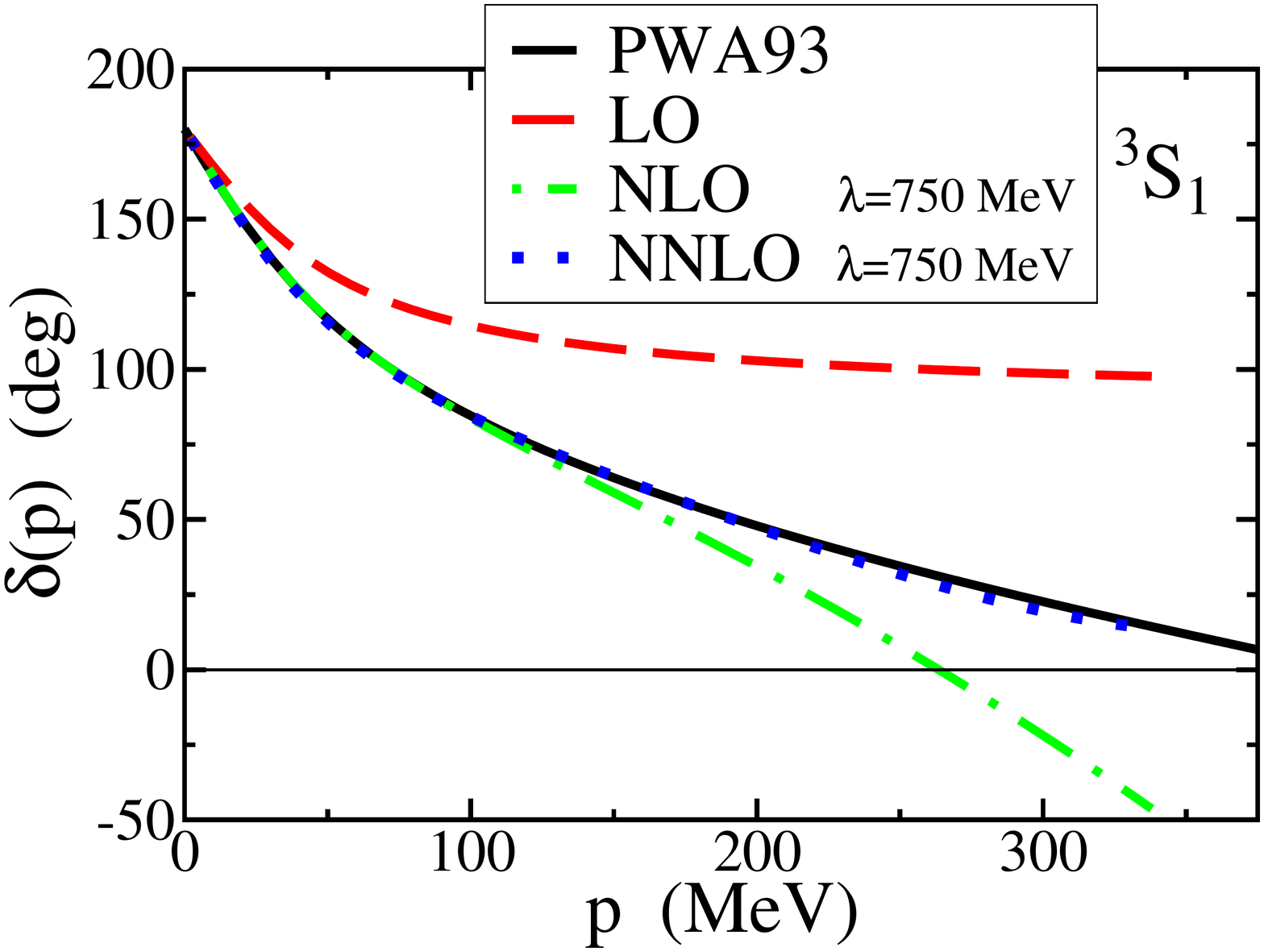}\includegraphics[width=6.1cm]{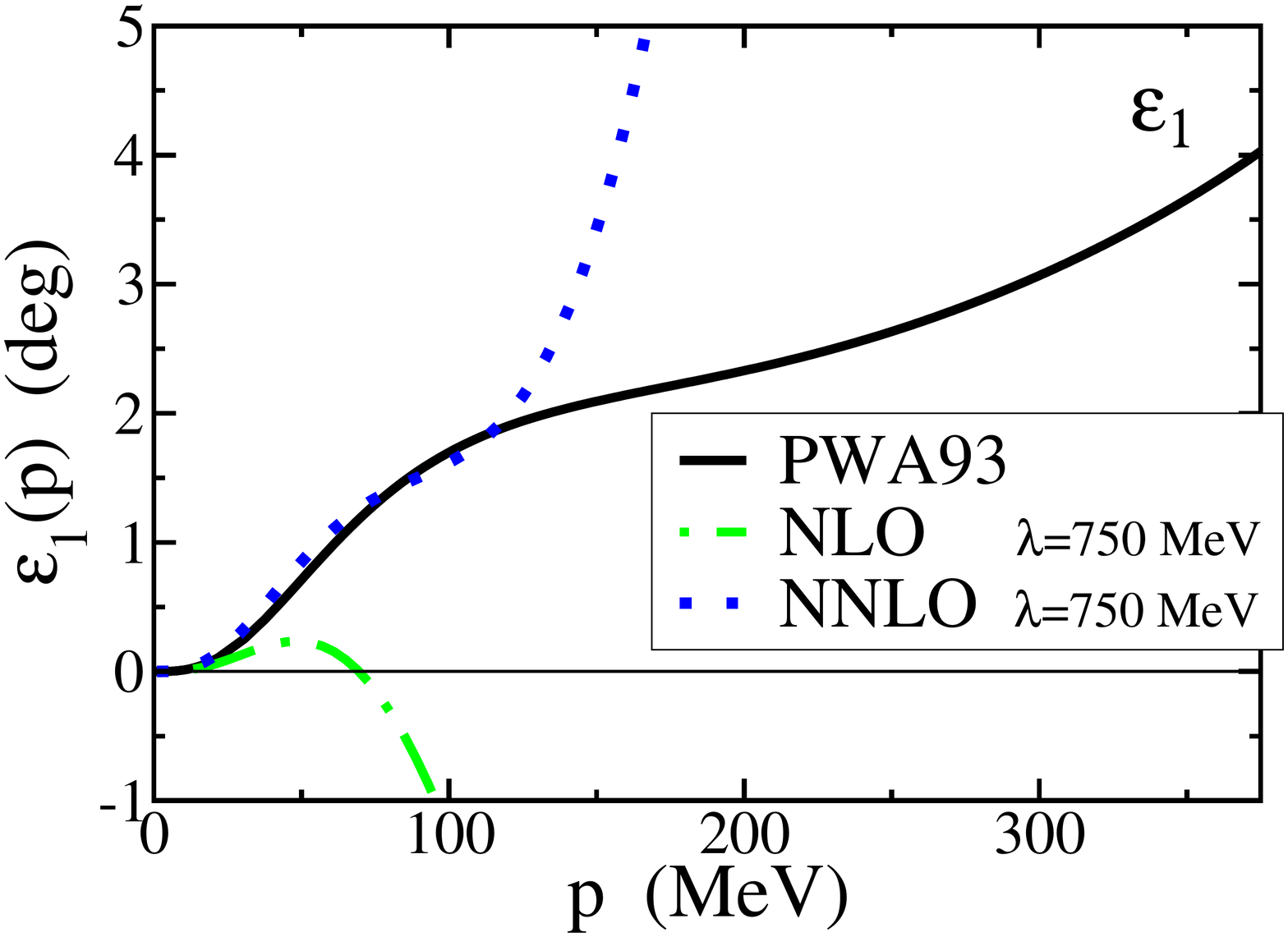}\includegraphics[width=6.1cm]{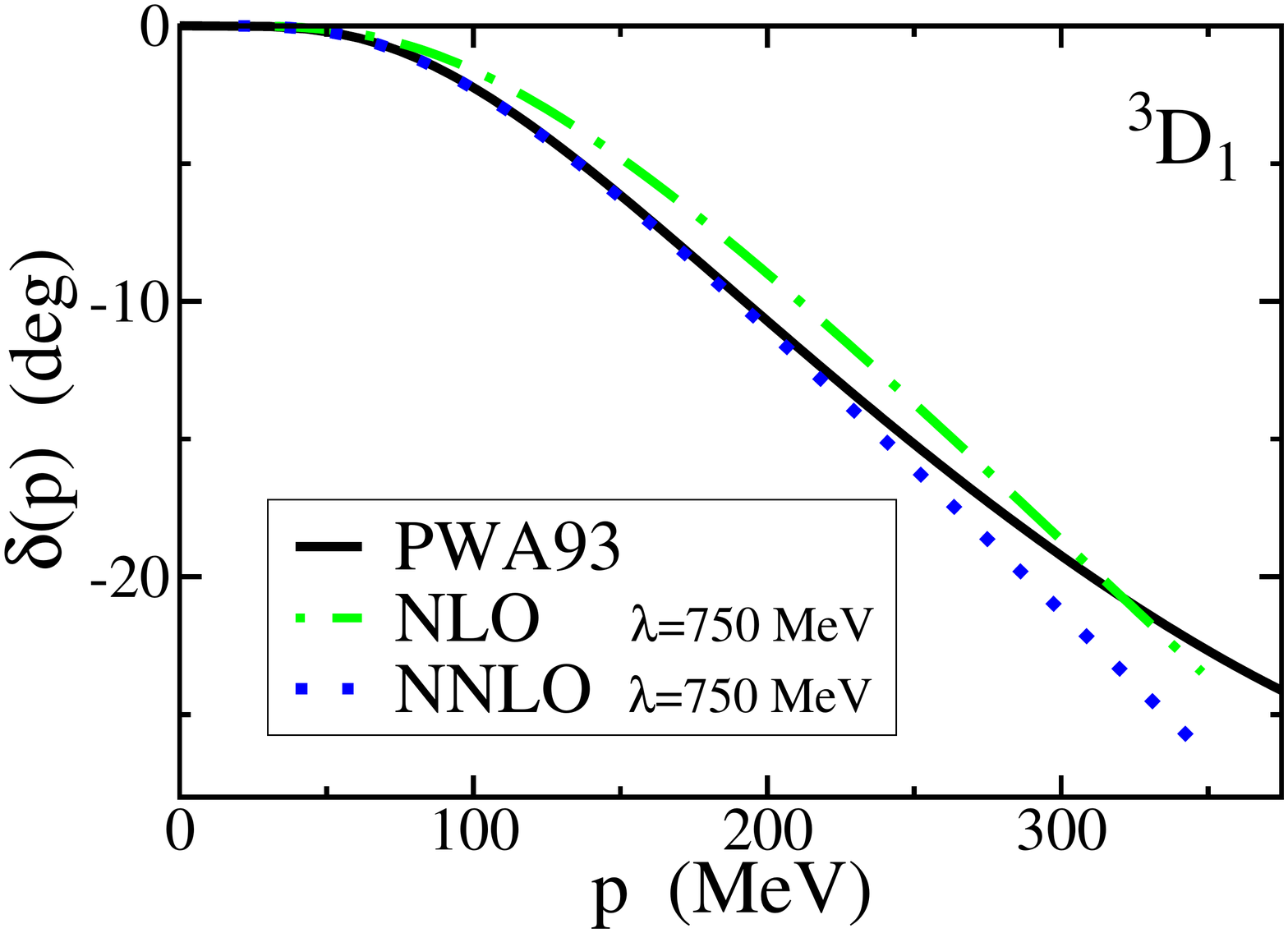}}

\label{J1gen}

\caption[a]{New results for the ${}^3S_1$, ${}^3D_1$, and
$\bar\epsilon_1 $ phase shifts plotted versus momentum in the center
of mass frame to NNLO, compared with the Nijmegen PWA93 partial wave
analysis. }
\end{figure*}

\begin{figure*}[t]

\hbox{\includegraphics[width=6.1cm]{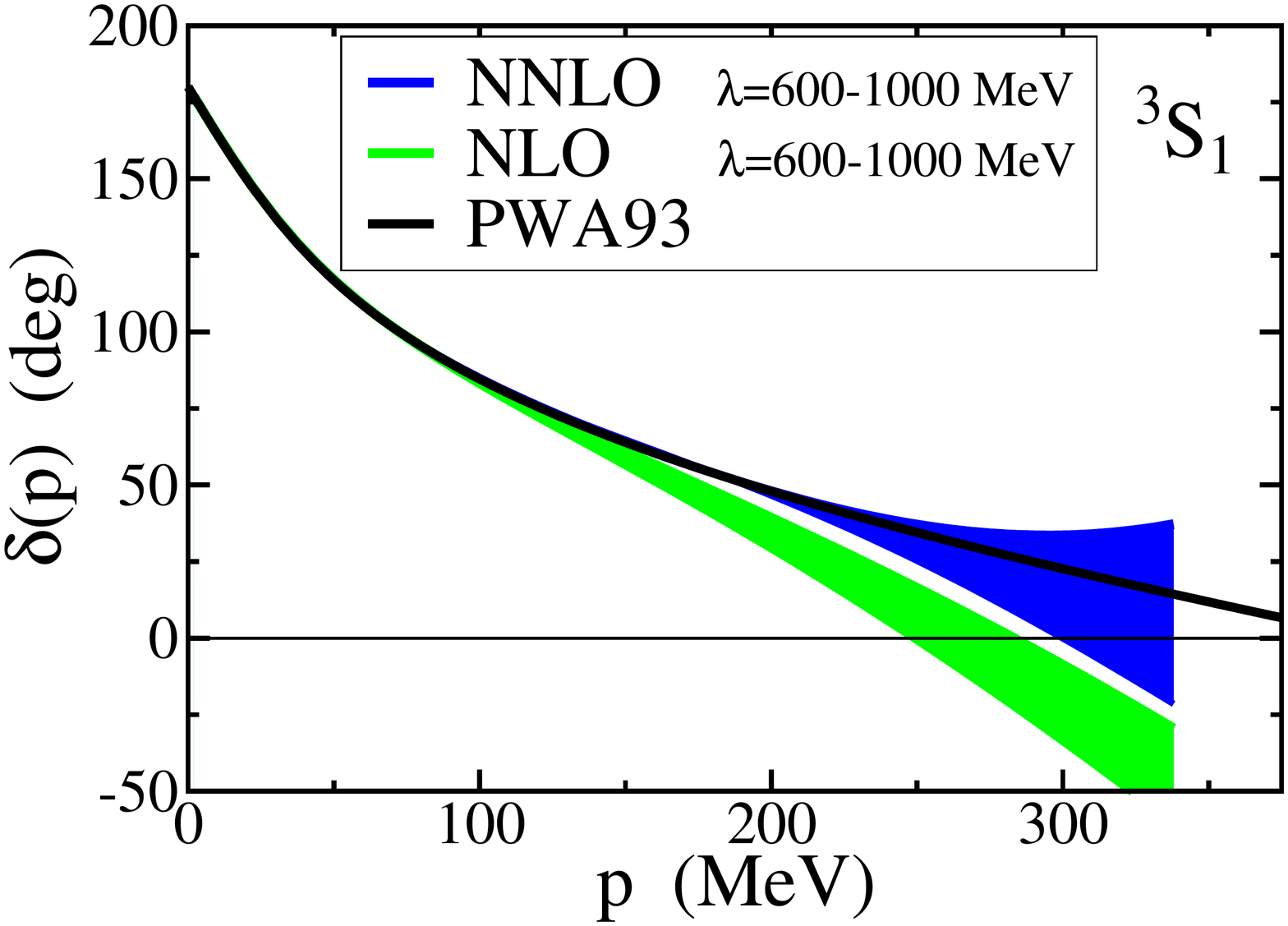}\includegraphics[width=6.1cm]{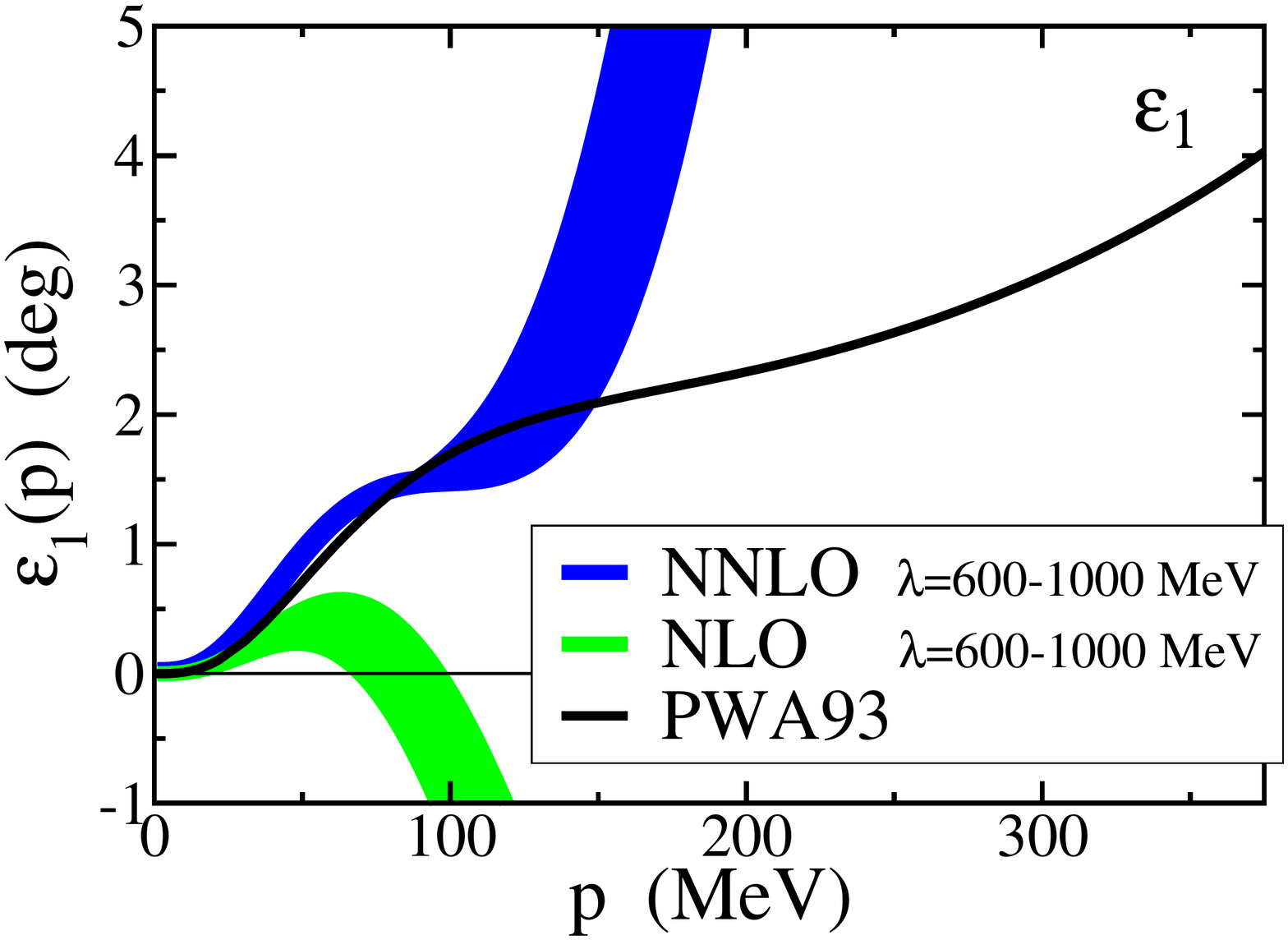}\includegraphics[width=6.1cm]{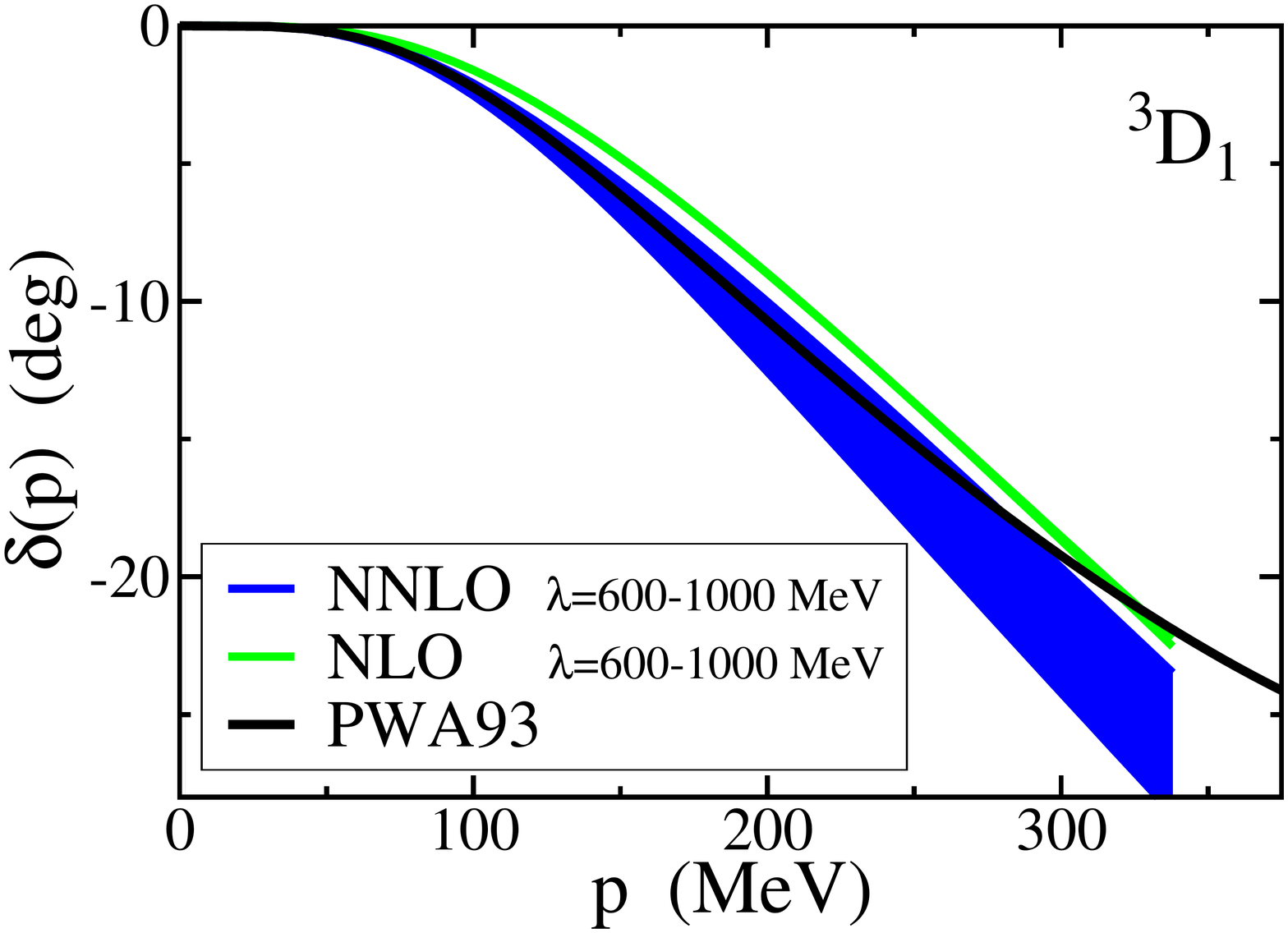}}

\label{J1var}

\caption[a]{The NLO (green band) and NNLO (blue band) results for the
${}^3S_1$, ${}^3D_1$, and $\bar\epsilon_1 $ phaseshifts showing their
variation as $\lambda$ is varied in the range $600 \MeV \le \lambda
\le 1000 \MeV$.} \end{figure*}

Our starting point is the assumption that the failure of the KSW
expansion is due to the singular short distance pion tensor
interaction, which can be eliminated by a shift in the contact
interactions of the EFT.  The underlying principles of EFT imply that
we are free to distort the short range pion interactions however we
please, as the counterterms serve to ensure the correct low energy
effects of short distance physics. We therefore choose the
modification in order to: (i) make it possible to analytically perform the
diagrammatic expansion; (ii) leave unaltered the KSW expansion of the
spin-singlet channel, since apparently no convergence
problem is encountered there. These considerations lead us to replace
the pion propagator $G_\pi(\bfq,m_\pi)$ by
 \beq 
 G_\pi(\bfq,m_\pi)
+ G_{(1,1)}(\bfq,\lambda) + G_{(1,0)}(\bfq,\lambda)
 \eqn{gmod} \eeq 
 where
the subscript $(I,J)$ indicates the isospin and spin of a fictitious
meson. Including couplings at the ends of the propagators, these
expressions are given by
 \beq G_\pi(\bfq,m_\pi) &=& { i \frac{g_A^2}{4
f_\pi^2}\frac{(\bfq\cdot\bfsigma_1)(\bfq\cdot\bfsigma_2)(\bftau_1\cdot\bftau_2)}{\bfq^2
+ m_\pi^2} }\cr G_{(1,0)}(\bfq,\lambda) 
&=& { i
\frac{g_A^2\lambda^2}{4f_\pi^2}\frac{(\bftau_1\cdot\bftau_2)}{\bfq^2 +
\lambda^2} }\ ,
 \eeq 
 and $G_{(1,1)}(\bfq,\lambda) =
-G_\pi(\bfq,\lambda)$. The $ G_{(1,1)} $ term looks like exchange of a
pion with the wrong sign propagator and mass $\lambda$, canceling the
short distance $1/r^3$ part of the pion-induced tensor interaction for
$r\lesssim 2\pi/\lambda$. The $G_{(1,0)}$ term is included to exactly
cancel $G_{(1,1)}$ (up to a contact interaction) in the
spin-singlet channel; it resembles the exchange of an $I=1$, $J=0$
meson, also of mass $\lambda$.  In the above expressions $g_A\simeq
1.25$ and $f_\pi\simeq 93\MeV$; ${\bfsigma}$ and $\bftau$ are spin and
isospin matrices respectively.  Note that the only free parameter is
the mass scale $\lambda$.  We expect that for $\lambda\gtrsim 2
\Lambda_{NN}$ the derivative expansion is not  adversely affected, and that the original KSW expansion is recovered  in the $\lambda\to\infty$ limit. 

We emphasize that we are not using $ G_{(1,1)} $ and $ G_{(1,0)}$ to
model real meson exchange, but only as a device to eliminate the
strong short distance behavior from the tensor pion exchange, putting
all that physics in the contact interactions which are fit to data.
Choosing the masses in $ G_{(1,1)} $ and $ G_{(1,0)}$ to both
equal $\lambda$ greatly simplifies the analytic computations.

\section{NNLO calculation of spin-triplet amplitudes}

Making use of the modified pion
propagator \eq{gmod} and classifying the mass scale $\lambda$ to also
be $O(Q)$, we have computed all the Feynman diagrams in \cite{Fleming:1999bs,Fleming:1999ee}
relevant for the ${}^3S_1$, ${}^3D_1$ and $\epsilon_1$ partial wave
channels. These diagrams are evaluated using dimensional regularization 
 and we choose the renormalization scale, $\mu=m_\pi$. The analytic formulas for our  NNLO calculations
will be given elsewhere; here we present the results graphically.  In
Fig.~2 we show our  results with $\lambda=750\MeV$ for the
${}^3S_1$, ${}^3D_1$ and $\epsilon_1$ phase shifts, compared with the
Nijmegen partial wave analysis \cite{Stoks:1993tb}. All three of our results are 
improvements over the NNLO KSW computation in \cite{Fleming:1999bs,Fleming:1999ee},
and with the exception of $\epsilon_1$, show signs of converging on
the correct answer.  The result for $\epsilon_1$ is less
convincing, but it should be noted that the anomalously small value
for $\epsilon_1$ in nature suggests that delicate cancellations are
at play, and one would only expect an EFT prediction to start
converging at high order in the expansion.


The dependence of our results on $\lambda$ is displayed in Fig.~3,
where the bands indicate the changes in the phase shifts over the
range $600\MeV\le \lambda\le 1000\MeV$ .  It is apparent from these
figures that our results are not extremely sensitive at low $p$ to the value we
take for $\lambda$.

The role of the scale $\lambda$ in these calculations can be easily
addressed given the analytic form we have derived for the scattering
amplitudes.  It may seem strange that $\lambda$---a regularization
scale---is being treated as $O(Q)$ which is our low energy expansion
scale.  In particular, one might worry that scattering amplitudes have
terms proportional to powers of $\lambda/\Lambda_{NN}$, which is
formally $O(Q)$ but numerically $>1$.  In fact though one can show
analytically that at each order in the expansion, contributions to the
amplitudes proportional to positive powers of $\lambda$ are all
absorbed into the counterterms available at that order.  Therefore the
amplitudes only depend on inverse powers of $\lambda$, and in the
$\lambda\to \infty$ limit the fictitious meson propagators in
\eq{gmod} decouple and one smoothly recovers the results of
\cite{Fleming:1999bs,Fleming:1999ee}.

\section{Discussion}

The EFT scheme we have presented here for computing NN scattering in
perturbation theory appears to converge well and preserve the
desirable feature of the KSW scheme that at each order the amplitude
can be computed as a well defined set of Feynman diagrams.  Unlike the
KSW scheme, there is now a new dimensionful parameter $\lambda$ which
regulates the short distance tensor interaction.  The manner with
which we have performed this regulation is certainly not unique, and
we have shown that our results are not particularly sensitive to the
value of $\lambda$, and that over a wide range for $\lambda$ the
variation of the phaseshifts are comparable to or smaller than higher
order corrections in the EFT expansion.

On the other hand, we know that by taking $\lambda\to \infty$ we
recover the KSW expansion, which fails to converge above $p\sim
100\MeV$.   The parameter $\lambda$ apparently plays a role
analogous to the renormalization scale $\mu$ in perturbative QCD.  The
scale $\mu$ is unphysical, and a nonperturbative QCD calculation will
not depend on it; however, at any finite order in perturbation
theory, amplitudes do depend on $\mu$, and varying $\mu$ corresponds
to reordering the perturbative expansion.  Choosing $\mu$
appropriately (e.g., 
via
the BLM scale-setting prescription \cite{Brodsky:1982gc}) can optimize
the perturbative expansion, while non-optimal choices for $\mu$ lead to poor convergence.  Similarly, $\lambda$ is an unphysical
parameter, and varying $\lambda$ constitutes a reordering of the the
EFT expansion, with smaller $\lambda$ resulting in more of the pion
interaction being accounted for in the resummed contact interactions.
Taking $\lambda \simeq 750\MeV$ appears to optimize the expansion,
while choosing $\lambda=\infty$ yields the standard KSW expansion
which fails to converge at relatively low momenta.

It is not possible to directly compare our expansion with the Weinberg
expansion results at a given order, since the calculations are
arranged differently. For example, one-pion and two-pion exchange appear at NLO and N${}^3$LO in the KSW expansion respectively, while they appear at LO and NLO  in the Weinberg expansion.  Nevertheless,
numerically our NNLO results compare favorably with the NLO Weinberg
expansion results in \cite{Epelbaum:1999dj}, with the exception of $\epsilon_1$ which is comparable to LO.  

In a subsequent paper we will present the detailed form of the
 amplitudes at NNLO for the partial waves presented here, as well as
 others.  Our theory provides a well defined prescription for
 computing a number of additional processes to NNLO, such as
 electromagnetic effects, including form factors, Compton scattering,
 polarizabilities, and radiative capture, and it will be interesting  to
 compare such results with experiment to thoroughly judge the efficacy
 of this method.

\section*{Acknowledgments}

We thank U. van Kolck and M. J. Savage for useful communications.
SRB is partly supported by NSF CAREER Grant No. PHY-0645570. DBK would
like to thank the Instituto de Fisica T\'eorica of the Universidad
Aut\'onoma de Madrid for hospitality during part of this project.  DBK
was supported in part by the DOE under contract DE-FGO3-00ER41132, by
the Spanish MEC grant SAB2006-0089 and project FPA2006-05423, and the
regional «Comunidad de Madrid« HEPHACOS project. AV was supported in
part by the Austrian Science Foundation, FWF, project No. M1006, as
well as the Sofja Kovalevskaja Award of the Humboldt foundation.


\bibliography{EFT_silas_121808}
\bibliographystyle{apsrev}

\end{document}